# *On The Reconstruction of Interaction Networks with Applications to Transcriptional Regulation*

*Adam A. Margolin*[1,2], *Ilya Nemenman*[2], *Chris Wiggins*[3], *Gustavo Stolovitzky*[4], *Andrea Califano*[1,2,5]

[1]Department of Biomedical Informatics, [2]Joint Centers for Systems Biology, [3]Department of Applied Physics and Applied Mathematics, [5]Institute for Cancer Genetics, Columbia University, New York, NY 10032
[4]IBM Computational Biology Center, IBM T.J. Watson Research Center, Yorktown Heights, N.Y. 10598

Genome-wide clustering of gene expression profiles [1] provides an important first step in studies of transcriptional regulatory networks. However, the organization of genes into co-regulated clusters is too coarse a representation to identify individual interactions. This is because as biochemical signals travel through cellular networks the expression of many genes that interact only indirectly may become strongly correlated. More generally, as has been appreciated in statistical physics, a long range order (that is, a high correlation among indirectly interacting random variables) can easily result from only short range, pairwise interactions [2]. Thus correlations, or *any other* local dependency measure, cannot be used as the only tool for the reconstruction of interaction networks.

Within the last years a number of sophisticated approaches for the reverse engineering of cellular networks (also called deconvolution) from gene expressions have emerged [3-5]. Their goal is to produce a high-fidelity representation of the cellular network topology as a graph, where genes are represented as nodes and direct regulatory interactions as edges. However, all approaches suffer to various degrees from problems such as overfitting, high computational complexity, reliance on non-realistic network models, or a critical dependency on supplementary data, available only for simple organisms. These limitations have relegated the successful application of most methods to simpler organisms, such as the yeast *S. cerevisiae*. Here we introduce *ARACNE* (Algorithm for the Reconstruction of Accurate Cellular Networks), a novel information-theoretic algorithm for the reverse-engineering of transcriptional networks from microarray data that overcomes some of these critical limitations. ARACNE compares favorably with existing methods and scales successfully to large network sizes. It is also general enough to deal with a variety of other network reconstruction problems.

Theoretical Background: We start by noting that with little temporal gene expression data available for higher eukaryotes, one is forced to study steady-state inter-gene statistical dependences only. We define these following the definition of [6], which builds on ideas from the Markov networks literature [7]. Briefly, by analogy with statistical physics, we write the joint probability distribution (JPD) of the stationary expressions of all genes, $P(\{g_i\})$, $i = 1,\ldots,N$, as:

$$P(\{g_i\}) = \frac{1}{Z}\exp\left[-\sum_i \phi_i(g_i) - \sum_{i,j}\phi_{ij}(g_i,g_j) - \sum_{i,j,k}\phi_{ijk}(g_i,g_j,g_k) - \cdots\right] \equiv \exp[-H(\{g_i\})] \qquad (1)$$

where $N$ is the number of genes, $Z$ is the *partition function*, $\phi_i(g_i)$ are *potentials*, and $H(\{g_i\})$ is the *Hamiltonian* that defines the system's statistics. The expansion in Eq. (1) does not define the potentials uniquely, and additional constraints of the Maximum Entropy type are needed to resolve the ambiguity [6]. Within such a model, a set of variables interacts *iff* the single potential that depends exclusively on these variables is nonzero. ARACNE aims precisely at identifying such potentials and eliminating the others even though their corresponding marginal JPDs may not factorize.

Since typical microarray sample sizes, $M$, are relatively small, inferring the exponential number of potential *n*-way interactions of Eq. (1) is infeasible and a set of simplifying assumptions must be made about the dependency structure. Eq. (1) provides a principled and controlled way to introduce such approximations. The simplest model is one where genes are assumed independent, i.e., $H(\{g_i\}) = \sum \phi_i(\{g_i\})$, such that the first-order potentials can be evaluated from the marginal probabilities, $P(g_i)$, which are in turn estimated from samples. As more data become available, we should be able to reliably estimate higher order marginals and incorporate the corresponding potentials progressively, such that for $M \to \infty$ the complete form of the JPD is restored. In fact, $M > 100$ is generally sufficient to estimate 2-way marginals in genomics problems, while $P(g_i, g_j, g_k)$ requires about an order of magnitude more samples. Thus we truncate Eq. (1) at the pairwise interactions only, $H(\{g_i\}) = \sum_i \phi_i(g_i) + \sum_{ij}\phi_{ij}(g_{ij})$. Within this approximation, all genes for which $\phi_{ij} = 0$ are declared non-interacting. This includes genes that are statistically independent (i.e., $P(g_i, g_j) \approx P(g_i)P(g_j)$), as well as genes that do not interact directly but are statistically dependent due to their interaction with others (i.e., $P(g_i, g_j) \neq P(g_i)P(g_j)$, but $\phi_{ij} = 0$). Since the number of potential pairwise interactions is quadratic in *N*, discriminating the latter situation is a formidable challenge for all network reconstruction algorithms that rely on statistical associations. However, under certain biologically realistic assumptions about the network topology, ARACNE provides a framework to reconstruct two-way interaction networks reliably from a finite number of samples in a computationally feasible time.

This formulation is reminiscent of spin glasses on random networks [8], particularly if the $g_i$ are binary. In this case, the genes are the Ising spins, and truncations to the first, second, or the third order potentials are steps towards the mean field, Bethe, and Kikuchi variational approximations [9-11].

The Algorithm: Within the assumption of a two-way network, all statistical dependencies can be inferred from pairwise marginals, and no higher order analysis is needed. Thus we identify candidate interactions by estimating pairwise gene-gene mutual information (MI), $I(g_i, g_j) \equiv I_{ij}$, an information-theoretic measure of relatedness that is zero *iff* $P(g_i, g_j) = P(g_i)P(g_j)$. We then filter MIs using a threshold, $I_0$, computed for a specific p-value, $p_0$, in the null-hypothesis of independent genes. This step is basically equivalent to the Relevance Networks [12] and suffers from the same critical limitations. In particular, genes separated by intermediaries may be co-regulated without implying physical interactions.

Thus, in its *second step*, ARACNE removes the vast majority of indirect candidate interactions using a well-known property of mutual information – the data processing inequality (DPI) [13] -- that has not been previously applied to the reverse engineering of networks. The DPI states that if genes $g_1$ and $g_3$ interact only through a third gene, $g_2$, (i.e., if the interaction network is $g_1 \leftrightarrow ... \leftrightarrow g_2 \leftrightarrow ... \leftrightarrow g_3$, and no alternative path exists between $g_1$ and $g_3$), then

$$I(g_1, g_3) \leq \min\left[I(g_1, g_2); I(g_2, g_3)\right]. \qquad (2)$$

Correspondingly, ARACNE starts with a network graph where each $I_{ij} > I_0$ is represented by an edge (*ij*). It then examines each gene triplet, for which all three MIs are greater than $I_0$, and removes the edge with the smallest value. Each triplet is analyzed irrespective of whether one of its edges has been marked for removal by a prior DPI application to a different triplet. Thus the network reconstructed by the algorithm is independent of the order in which the triplets are examined.

Theorem 1. If MIs can be estimated with no errors, then ARACNE reconstructs the underlying interaction network exactly, provided this network is a tree and has only pairwise interactions.

*Proof of Theorem 1.* First, notice that for every pair of nodes *i* and *k* not connected by a true direct interaction there is at least one other node *j* that separates them on the network tree. Applying the DPI to the (*ijk*) triplet leads to removal of the (*ik*) edge. Thus only true edges survive. Similarly, every removed edge is not present in the true network. Consider some (*ijk*) triplet. One of these nodes, say *j*, may separate the other two. In this case the removed edge (*ik*) is clearly not in the true tree. Alternatively, there may be no separating node, and one may be able to move between any genes in the triplet without going through the third one. In this case none of the three edges is in the true graph, and any edge DPI removes is fictitious. Thus all removed edges are indirect, while all remaining edges are factual. The network is reconstructed exactly.

Unlike standard spanning tree reconstruction methods (e.g. Chow and Liu [14]), ARACNE is not limited to trees. In fact,

Theorem 2. Chow-Liu (CL) maximum mutual information tree is a subnetwork of the network reconstructed by ARACNE.

*Proof of Theorem 2.* We notice that, without a loss of generality, we can assume that the Chow-Liu tree and the ARACNE construction span all the nodes of the network. If this is not the case, that is, a few connected clusters exist (separated by edges with zero MI), then for the purpose of this theorem we can complete CL and ARACNE structures by the same edges with zero MI without formation of additional loops, till they become spanning. Now suppose that the theorem is false and there exists an edge *(ij)* that belongs to the (completed) CL tree, but does not belong to the ARACNE reconstruction. Since the CL construct is a tree, this edge separates it into two separate trees $T_i$ and $T_j$ that contain the *i*'th and the *j*'th nodes respectively. Since ARACNE has removed the *(ij)* link, there exists a node *k*, for which $\min(I_{ik}, I_{jk}) > I_{ij}$. Without a loss of generality, let *k* be in $T_i$. Then replacing the *(ij)* edge in the Chow-Liu tree by the *(jk)* edge will form no loops and will preserve the tree structure. This will increase the total MI of the CL reconstruction by $I_{jk} - I_{ij} > 0$. Thus the original tree is not the maximum MI tree. We arrive at a contradiction, which proves the theorem.

Theorem 3. Let $\pi_{ik}$ be the set of nodes forming the shortest path in the network between nodes *i* and *k*. Then, if MIs can be estimated without errors, ARACNE reconstructs an interaction network without false positives edges, provided: (a) the network consists only of pairwise interactions, (b) for each $j \in \pi_{ik}$, $I_{ij} \geq I_{ik}$. Further, ARACNE does not produce any false negatives, and the network reconstruction is exact *iff* (c) for each directly connected pair *(ij)* and for any other node *k*, we have $I_{ij} \geq \min(I_{jk}, I_{ik})$.

*Proof of Theorem 3.* To prove the absence of false positives, we notice that, for every candidate edge *(ik)* that is not actually in the network, there is at least one node *j*, such that $j \in \pi_{ik}$. Applying DPI to the *(ijk)* triplet will remove the *(ik)* edge. Further, we notice that if (c) is satisfied, then any application of DPI will not remove a true edge. However, if (c) does not hold, a true edge will be removed. This completes the proof.

Note that tree networks satisfy all conditions of Theorem 3, while loopy topologies may or may not. In particular, networks with three-gene loops definitely violate (c) [but may still satisfy (a) and (b)], and *every* such loop will be opened along the weakest edge. For a tree, there is a unique path that connects two nodes. Similarly, for networks that satisfy (a) and (b), the shortest path dominates inter-node information transfer. We call these networks *locally tree-like*. In other words, an interaction is retained by ARACNE if and only if there exist no alternate paths, via one or more intermediaries or branches on the network graph, which are a better explanation for the information exchange between two genes. Since biochemical dynamics is inherently stochastic, statistical interactions over more than a few separating edges are generically weak. Thus we believe that the local tree assumption is biologically realistic (a notable exception is the feed forward loop, found to be over-represented in biological circuits [15]), and we expect ARACNE to produce low false positive rates in practice, while having a minimal impact on false negative ones. We will demonstrate this using a synthetic dataset below.

In the current implementation of the algorithm, we use a computationally efficient Gaussian Kernel estimator [16] of MI. Given two-dimensional samples, $\vec{z}_i \equiv \{x_i, y_i\}, i = 1 \ldots M$, the JPD is approximated as $f(\vec{z}) = 1/M \sum_i h^{-2} G\left(h^{-1} |\vec{z} - \vec{z}_i|\right)$, where $G(\ldots)$ is the bivariate standard normal density. With $f(x)$ and $f(y)$ being the marginals of $f(\vec{z})$, the MI is:

$$I(\{x_i\}, \{y_i\}) = \frac{1}{M} \sum_i \log \frac{f(x_i, y_i)}{f(x_i) f(y_i)} \qquad (3)$$

Since MI is reparameterization invariant, we copula-transform *x* and *y* for MI estimation. This decreases the influence of arbitrary transformations involved in microarray data preprocessing and removes the need to consider position-dependent kernel width, *h*, which might be needed for the original, non-uniform, data. This estimator is asymptotically unbiased for $M \to \infty$, as long as $h(M) \to 0$ and $[h(M)]^2 M \to \infty$. However, for finite *M*, the bias strongly depends on $h(M)$, and the correct choice is not universal. Fortunately, ARACNE's performance does not depend directly on the accuracy of the MI estimate, but rather on the accuracy of the estimation of MI ranks: to test if MI is statistically significant or to apply DPI, one only needs to check if $I_{ij} > I_0$, or if $I_{ij} > I_{ik}$, respectively; that is, only to rank MI estimates. It turns out that for fixed *h* the bias tends to cancel out, especially for $\bar{I}_{ij} \approx \bar{I}_{kl}$, and the ordering of MI estimates is only weakly dependent on *h* and is stable even when MI itself is uncertain. Thus selecting a single "ensemble best" value of *h* rather than searching for the best kernel width for each estimate (a computationally intensive operation) impedes performance very little. With such a choice, ARACNE's complexity is $O(N^3 + N^2 M^2)$, where *M* is the number of samples, and *N* is the number of genes. This is low enough to effectively analyze networks with tens of thousands of genes. We refer the reader to [17] for details of selection of the kernel width as well as the other adjustable parameter, the DPI tolerance, $\tau$, which can be used to further minimize the impact of potential MI estimation errors by transforming DPI inequalities to the form $I_{ij} \leq I_{ij}(1 - \tau)$.

Performance: We analyzed ARACNE's performance on reconstructing synthetic networks proposed by [18] specifically as a benchmark for reverse engineering algorithms ([17, 19] also present applications to the human B cell network). The networks consist of 100 genes and 200 interactions organized in an Erdös-Rényi (random) [20] or a scale-free [21] topology, and they evolve according to a multiplicative Hill dynamics. Such networks present a formidable challenge to reconstruction algorithms due to (a) their realistic complexity, (b) the presence of many regulatory loops, (c) the presence of a few highly interconnected genes (for the scale-free version), and (d) the biologically motivated non-linear transcriptional dependencies among genes. To generate synthetic microarrays, we randomly vary the efficiency of gene synthesis and degradation reactions for each synthetic sample at the beginning of each simulation. This models the sampling of a population of distinct cellular phenotypes at random time points (but in equilibrium).

ARACNE's performance is compared against Relevance Networks (RNs) [12] and Bayesian Networks (BNs) [7], as implemented by [22]. RNs, which are equivalent to ARACNE without the DPI step, are important to characterize the improvement associated with the introduction of the DPI, while BNs have emerged as some of the best available reverse engineering methods and provide an ideal comparative benchmark. The benchmark measures are *recall,* $N_{TP} / (N_{TP} + N_{FN})$, and precision, $N_{TP}/(N_{TP}+N_{FP})$, which, respectively, measure the fraction of true interactions correctly inferred by the algorithm and the fraction of genuine interactions among all predicted ($N_{TP}$, $N_{FP}$, $N_{TN}$, and $N_{FN}$ stand for true/false positives/negatives). Precision vs. Recall curves (PRCs) are a better match than the more familiar ROC curves for problems where $N_{TN}$ is far greater than $N_{TP}$, which is the case in large sparse networks.

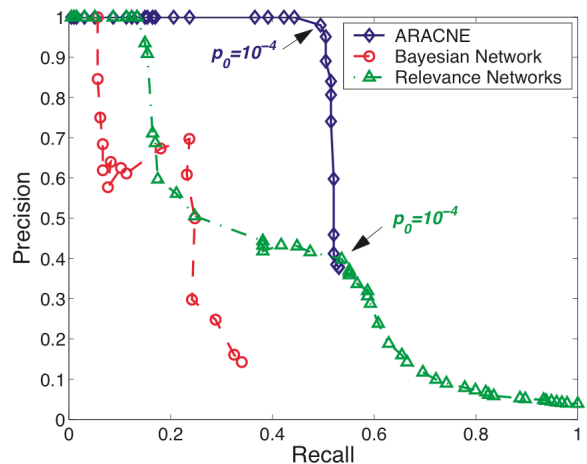

PRCs are shown in the Figure for all three comparative algorithms. We varied the MI threshold and the Dirichlet pseudocount to generate the PRCs for ARACNE/RNs and BNs respectively. ARACNE performs consistently better than BNs and RNs for both types of topologies considered. That is, for any reasonable precision (i.e. $>40\%$), ARACNE has a significantly higher recall than the other methods, and its precision reaches ~100% at significant recall values. Such high precision is necessary to guide experimental validation of the method's predictions. Using 1,000 samples, for both topologies, over half of all edges can be inferred with hardly any false positives. Further, performance degrades gracefully as the sample size decreases and is highly stable with respect to the choice of the kernel width (cf. [17]). The black arrows on the Figure indicate the a priori optimal operating points for the algorithms, where only O(1) false positives are expected.

Summary: ARACNE appears (a) to achieve very high precision and substantial recall, (b) to be stable with respect to the choice of parameters, and (c) to achieve substantial recall and high precision even with very few data points (125). ARACNE drastically improves network inference due to its efficiency in filtering false-positives, although it may potentially open up some loops of interacting genes and it neglects higher order interactions. In [17] we address these issues and offer suggestions for future investigation.